\documentclass[12pt]{iopart}
\usepackage{epsfig}
\usepackage{citesort}
\begin{document}
\newcommand{\no}{\nonumber\\}
\def\tgb{\mbox{$\tan{\beta}~$}}
\def\bsg{$b\to s \gamma$~}
\def\eps{$\varepsilon$~}
\def\epspeps{$\varepsilon^{\prime}/\varepsilon$~}
\def\Lsoft{${\cal L}_{SB}$~}
\def\mch{$m_{\chi^{\pm}}$~}
\def\mneu{$m_{\chi^{0}}$~}
\def\mglu{$m_{\tilde{g}}$~}
\def\stop{$m_{\tilde{t}}$~}
\def\mgrav{$m_{3/2}$~}
\def\Ibanez{Iba\~{n}ez~}
\def\Munoz{Mu\~{n}oz~}
\def\al{\alpha}
\def\be{\beta} 
\newcommand{\BXcenu}{B\rightarrow X_c e \nu}
\newcommand{\mub}{\mu_b}
\newcommand{\mb}{m_b}
\newcommand{\alphas}{\alpha_s}
\newcommand{\alphae}{\alpha_e}
\newcommand{\BRg}{{\rm BR}(B\to X_s \gamma)}
\newcommand{\BR}{{\rm BR}}
\newcommand{\Bsg}{B\to X_s \gamma}

\def\be{\begin{equation}}
\def\ee{\end{equation}}
\def\bea{\begin{eqnarray}}  
\def\eea{\end{eqnarray}}   
\def\etal{{\it et al.}}   
\def\eg{{\it e.g.}}
\def\ie{{\it i.e.}}
\def\Frac#1#2{\frac{\displaystyle{#1}}{\displaystyle{#2}}}
\def\lsim{\raise0.3ex\hbox{$\;<$\kern-0.75em\raise-1.1ex\hbox{$\sim\;$}}}
\def\gsim{\raise0.3ex\hbox{$\;>$\kern-0.75em\raise-1.1ex\hbox{$\sim\;$}}}
\renewcommand{\O}{{\cal O}}
\def\ap#1#2#3{     {\it Ann. Phys. (NY) }{\bf #1} (#2) #3}
\def\arnps#1#2#3{  {\it Ann. Rev. Nucl. Part. Sci. }{\bf #1} (#2) #3}
\def\npb#1#2#3{    {\it Nucl. Phys. }{\bf B #1} (#2) #3}
\def\npbps#1#2#3{    {\it Nucl. Phys. }(Proc. Suppl.){\bf B #1} (#2) #3}
\def\plb#1#2#3{    {\it Phys. Lett. }{\bf B #1} (#2) #3}
\def\prd#1#2#3{    {\it Phys. Rev. }{\bf D #1} (#2) #3}
\def\prep#1#2#3{   {\it Phys. Rep. }{\bf #1} (#2) #3}
\def\prl#1#2#3{    {\it Phys. Rev. Lett. }{\bf #1} (#2) #3}
\def\ptp#1#2#3{    {\it Prog. Theor. Phys. }{\bf #1} (#2) #3}
\def\rmp#1#2#3{    {\it Rev. Mod. Phys. }{\bf #1} (#2) #3}
\def\zpc#1#2#3{    {\it Zeit. f\"ur Physik }{\bf C #1} (#2) #3}
\def\mpla#1#2#3{   {\it Mod. Phys. Lett. }{\bf A #1} (#2) #3}
\def\sjnp#1#2#3{   {\it Sov. J. Nucl. Phys. }{\bf #1} (#2) #3}
\def\yf#1#2#3{     {\it Yad. Fiz. }{\bf #1} (#2) #3}
\def\nc#1#2#3{     {\it Nuovo Cim. }{\bf #1} (#2) #3}
\def\jetpl#1#2#3{  {\it JETP Lett. }{\bf #1} (#2) #3}
\def\ibid#1#2#3{   {\it ibid. }{\bf #1} (#2) #3}

\title[]{FCNC in SUSY models with non-universal A-terms}

\author{Shaaban Khalil}

\address{Centre for Theoretical Physics, University of Sussex, Brighton 
BN1 9QJ,~~~U.~K.}  

\address{Ain Shams University, Faculty of Science, Cairo, 11566, Egypt.}

\begin{abstract}
We study the inclusive branching ratio for $\Bsg$
in a class of string-inspired SUSY models with
non--universal soft-breaking $A$--terms.
We show that \bsg do not severely constrain the non--universality
of these models and the parameter regions which are important for generating  
sizeable contribution to \epspeps, of order $2 \times 10^{-3}$, are not
excluded. We also show that the CP asymmetry of this decay is predicted to be
much larger than the standard model prediction in a wide region of the
parameter space. In particular, it can be of order $10-15\%$ which can be
accessible at $B$ factories.
\end{abstract}


\section{Introduction}

The inclusive radiative decay  $\Bsg$ is known to provide a valuable constraint on 
any new physics beyond the standard model (SM).
The most recent result reported by CLEO collaboration
for the total (inclusive) B meson branching ratio $\Bsg$ is~ \cite{CLEO}
\be
{\rm BR}(B\to X_s\gamma)=(3.15\pm 0.35\pm 0.32\pm 0.26)\times 10^{-4}
\label{bsgCLEO1}
\ee
where the first error is statistical, the second systematic, and the
third one accounts for model dependence. From this result the following
bounds (each of them at 95\% C.L.) are obtained
\be
2.0\times 10^{-4} <
{\rm BR}(B\to X_s\gamma)< 4.5 \times 10^{-4}.
\label{bsgCLEO2}
\ee
In addition the ALEPH collaboration at LEP reported a compatible measurement
of the corresponding branching ratio for b hadrons at the Z resonance
\cite{ALEPH}.

It is well known that these experimental limits on 
\bsg cause a dramatic reduction of the allowed
parameter space in case of universal soft terms \cite{bsgSUSY,BBMR}. 
However, it has been emphasized recently that the non-degenerate
$A$--terms can generate the experimentally observed CP violation $\varepsilon$ and
$\varepsilon'/\varepsilon$ even with a vanishing
$\delta_{CKM}$~\cite{barr,khalil1,khalil2,vives}, i.e., fully supersymmetric CP violation
in the kaon system is possible in a class of models with non--universal
$A$--terms. So one may worry if these constraints would be even more severe in the case 
of the non-degenerate $A$--terms.
The non--universal $A$--terms could enhance the gluino contribution to \bsg decay 
which is usually very small in the universal case, being
proportional to the mass insertions $(\delta_{LR}^d)_{23}$. It could also give large 
contributions to the chargino amplitude through the $(\delta_{LR}^u)_{23}$.
Therefore a careful analysis of the \bsg predictions, including
the full SUSY contributions, is necessary in this scenario.

In most of analysis  universal or degenerate $A$-terms have
been assumed, i.e., $(A_{U,D,L})_{ij}=A$ or $(A_{U,D,L})_{ij}=A_{U,D,L}$.
This is certainly a nice simplifying assumption,
but it removes some interesting degrees of freedom.
For example, every $A$-term would, in general have an independent CP
phase, and in principle we would have $27(=3 \times 3 \times 3)$ independent
CP phases. However, in the universal assumption only one independent CP phase
is allowed. The situation drastically changes if we are to allow for non-degenerate
{A} terms with different and independent CP phases.
For example, the off-diagonal element of the squark (mass)$^2$ matrix,
say $(M_Q^2)_{12}$, includes the term proportinal to
$(A_U)_{1i}(A_U^{\dag})_{i2}$. However, in the universal or the degenerate case this term 
is always real. Furthermore, these off-diagonal elements play an important
role in \eps and \epspeps~\cite{khalil1,khalil2,vives}.

The major bulk of this talk will be devoted to the discussion of the \bsg constraints for the 
SUSY models with non--universal $A$--terms studied in Refs.~\cite{khalil1,khalil2,vives} 
following the work down in Ref.~\cite{emidio}.
We will also mention to the effect of the flavour--dependent phases of the $A$--terms on 
the CP asymmetry in the inclusive $\Bsg$ decay~\cite{david}.

\section{String inspired models with non-degenerate  $A$--terms}
In this work we consider the class of string inspired model which
has been recently studied in
Refs.~\cite{khalil1,khalil2,vives}. In
this class of models, the trilinear $A$--terms of the soft SUSY
breaking are non--universal. It was shown that this
non--universality among the $A$--terms plays an important role on
CP violating processes. In particular, it has been shown that
non-degenerate $A$-parameters can generate the experimentally
observed CP violation $\varepsilon$ and $\varepsilon'/\varepsilon$
even with a vanishing $\delta_{\mathrm{CKM}}$.

Here we consider two models for non-degenerate $A$--terms. The
first model (model A) is based on weakly coupled heterotic strings,
where the dilaton and the moduli fields contribute to SUSY
breaking~\cite{ibanez1}. The second model (model B) is based on type
I string theory where the gauge group $SU(3) \times U(1)_Y$ is
originated from the $9$ brane and the gauge group $SU(2)$ is
originated from one of the $5$ branes~\cite{ibanez2}.

\subsection{Model A}
We start with the weakly coupled string-inspired supergravity
theory.
In this class of models, it is assumed that the superpotential of
the dilaton ($S$) and moduli ($T$) fields is
generated by some non-perturbative mechanism and   
the $F$-terms of $S$ and $T$ contribute to the SUSY breaking.
Then one can parametrize the $F$-terms as~\cite{ibanez1}
\be
F^S = \sqrt{3} m_{3/2} (S+S^*) \sin\theta,\hspace{0.75cm} F^T
=m_{3/2} (T+T^*) \cos\theta .
\ee
Here $m_{3/2}$ is the gravitino mass, $n_i$ is the modular weight
and $\tan \theta$ corresponds to the ratio between the $F$-terms of $S$
and $T$.
In this framework, the soft scalar masses $m_i$ and the gaugino masses
$M_a$ are given by~\cite{ibanez1}
\begin{eqnarray}
m^2_i &=& m^2_{3/2}(1 + n_i \cos^2\theta), \label{scalar}\\ M_a
&=& \sqrt{3} m_{3/2} \sin\theta .\label{gaugino}
\end{eqnarray}
The  $A^{u,d}$-terms are  written as
\begin{eqnarray}
(A^{u,d})_{ij} &=& - \sqrt{3} m_{3/2} \sin\theta- m_{3/2}
\cos\theta (3 + n_i + n_j + n_{H_{u,d}}), \label{trilinear}
\end{eqnarray}
where $n_{i,j,k}$ are the modular weights of the fields
that are coupled by this $A$--term. 
If we assign $n_i=-1$ for the third family and $n_i=-2$
for the first and second
families (we also assume that $n_{H_1}=-1$ and $n_{H_2}=-2$) we find the following 
texture for the $A$-parameter matrix at the string scale
\begin{equation}
A^{u,d} = \left (
\begin{array}{ccc}
x_{u,d} & x_{u,d} & y_{u,d}\\
x_{u,d} & x_{u,d} & y_{u,d} \\
y_{u,d} & y_{u,d} & z_{u,d}  
\end{array}
\right),
\label{AtermA}
\end{equation}
where
\begin{eqnarray}
x_u&=& m_{3/2}(-\sqrt{3} \sin\theta + 3  \cos\theta),\\
x_d&=&y_u= m_{3/2}(-\sqrt{3} \sin\theta + 2  \cos\theta),\\
y_d&=&z_u= m_{3/2}(-\sqrt{3} \sin\theta + \cos\theta),\\
z_d&=&-\sqrt{3}m_{3/2}\sin\theta.
\end{eqnarray}

The non--universality of this model is
parameterized by the angle $\theta$ and the value $\theta =\pi/2$
corresponds to the universal limit for the soft terms. In order to
avoid negative mass squared in the scalar masses we restrict  
ourselves to the case with $\cos^2 \theta < 1/2$. Such
restriction on $\theta$ makes the non--universality in the whole
soft SUSY breaking terms very limited. However, as shown in
\cite{khalil1,khalil2}, this small range of variation for the
non--universality is enough to generate sizeable SUSY CP violations
in K system.

\subsection{Model B}
This model is based on type I string theory and like model A, it is a good candidate for
generating sizeable SUSY CP violations. In type I string theory, non--universality 
in the scalar masses, $A$--terms and gaugino masses
can be naturally obtained~\cite{ibanez2}. Type I models contain
either 9 branes and three types of $5_i (i=1,2,3)$ branes or $7_i$
branes and 3 branes. From the phenomenological point of view there
is no difference between these two scenarios. Here we consider the
same model used in Ref.~\cite{vives}, where the gauge group
$SU(3)_C \times U(1)_Y$ is associated with 9 brane while $SU(2)_L$
is associated with $5_1$ brane.

If SUSY breaking is analysed, as in model A,
in terms of the vevs of the dilaton and moduli fields \cite{ibanez2}
\be
F^S = \sqrt{3} m_{3/2} (S+S^*) \sin\theta,\hspace{0.75cm} F^{T_i} 
=m_{3/2} (T_i+T_i^*) \Theta_i \cos\theta~,
\ee
where the angle $\theta$ and the parameter $\Theta_i$ with
$\sum_i \left|\Theta_i\right|^2=1$, just parametrize the direction of the
goldstino in the $S$ and $T_i$ fields space .
Within this framework, the gaugino masses are~\cite{ibanez2}
\bea
M_1 &=& M_3 = \sqrt{3} m_{3/2} \sin\theta ,\\ M_2 &=& \sqrt{3}
m_{3/2} \Theta_1 \cos \theta .\label{m2}
\label{gauginoB}
\eea
In this case the quark doublets and the Higgs fields are assigned to
the open string which spans between the $5_1$ and $9$ branes.
While the quark singlets correspond to the open string which starts  
and ends on the $9$ brane, such open string includes three sectors
which correspond to the three complex compact dimensions. If we
assign the quark singlets to different sectors we obtain
non--universal $A$--terms. It turns out that in this model the 
trilinear couplings $A^u$ and $A^d$ are given
by~\cite{ibanez2,vives}
\begin{equation}
A^u=A^d = \left (
\begin{array}{ccc}
x & y & z\\
x & y & z \\
x & y & z
\end{array}
\right),
\label{AtermB1}
\end{equation}
where
\begin{eqnarray}
x &=& - \sqrt{3} m_{3/2}\left(\sin\theta + (\Theta_1 - \Theta_3) \cos\theta
\right),\\
y &=& - \sqrt{3} m_{3/2}\left(\sin\theta + (\Theta_1 - \Theta_2) \cos\theta
\right),\\
z &=& - \sqrt{3} m_{3/2} \sin\theta.
\label{AtermB2}
\end{eqnarray}
The soft scalar masses for quark-doublets and Higgs fields
$(m^2_L)$, and the quark-singlets $(m^2_{R_i})$ are given by
\bea
m^2_L &=&  m_{3/2}^2 \left( 1- \frac{3}{2} (1-\Theta_1^2) \cos^2
\theta\right) ,\\ m^2_{R_i} &=&  m_{3/2}^2 \left( 1- 3 \Theta_i^2 \cos^2
\theta\right),  
\label{scalarB}
\eea
where $i$ refers to the three families.
For $\Theta_{i} = 1/\sqrt{3}$ the $A$--terms and the
scalar masses are universal while the gaugino masses could be
non--universal. The universal gaugino masses are obtained at   
$\theta=\pi/6$.

In models with non-degenerate $A$--terms we have to fix the Yukawa
matrices to completely specify the model. 
Here we assume that the Yukawa texture has the following form   
\be
\hspace{-1.5cm}Y^u=\frac{1}{v\cos{\beta}} {\rm diag}\left(
m_u,m_c,m_t\right)~,~~
Y^d=\frac{1}{v\sin{\beta}}  V^{\dagger} \cdot {\rm diag }
\left(m_d, m_s, m_b\right) \cdot  V
\label{Yuk2}
\ee

\section{\bsg constraints vs. non--universality}

Theoretical study of \bsg decay is given by the effective Hamiltonian
\be
H_{eff}=-\frac{4G_F}{\sqrt{2}}V_{32}^{\star}V_{33}\sum_{i=1}^{8}
C_i(\mu_b) Q_i(\mu_b)
\label{Heff}
\ee
where the complete basis of operators in the SM can be found in
Ref.~\cite{bsgNLO}. Recently the main theoretical uncertainties present in the previous 
leading order (LO) SM calculations have been reduced by including the
NLO corrections to the \bsg decay, through
the calculation of the three-loop anomalous dimension matrix of the
effective theory \cite{bsgNLO}.
The relevant SUSY contributions to the effective Hamiltonian in Eq.(\ref{Heff})
affect only the $Q_7$ and $Q_8$ operators,
the expression for these operators are given (in the usual notation) by
\bea Q_7&=&\frac{e}{16 \pi^2}m_b\left(
\bar{s}_L\sigma^{\mu\nu} b_R\right) F_{\mu\nu}~,
\no 
Q_8&=&\frac{g_s}{16
\pi^2}m_b\left( \bar{s}_L\sigma^{\mu\nu} T^a b_R\right) G^a_{\mu\nu}~.
\label{operators}
\eea
The Wilson coefficients $C_i(\mu)$ are evaluated at the renormalization scale
$\mu_b\simeq O(m_b)$  by including the
NLO corrections \cite{bsgNLO}.
They can be formally decomposed  as follows
\be
C_i(\mu)=C_i^{(0)}(\mu)+\frac{\alphas(\mu)}{4\pi}C_i^{(1)}(\mu) +
\O(\alphas^2).
\ee
where $C_i^{(0)}$ and $C_i^{(1)}$ stand for the LO and NLO order
respectively.
The SUSY contributions to the Wilson coefficients $C_{7,8}^{(0,1)}$
are obtained by calculating
the \bsg and $b\to s g$ amplitudes at EW scale respectively.
The LO contributions to these amplitudes are given by the 1-loop
magnetic-dipole and chromomagnetic dipole penguin diagrams respectively,
mediated by charged Higgs boson, chargino, gluino, and neutralino exchanges.
The corresponding results for these amplitudes can be found in Ref.\cite{BBMR}.
We point out that
the SUSY models with non--universal $A$--terms may induce non-negligible
contributions to the dipole operators $\tilde{Q}_{7,8}$
which have opposite chirality with respect to $Q_{7,8}$.
It is worth mentioning that
these operators are also induced in the SM and in the MSSM with supergravity
scenario, but their contributions
are negligible being suppressed by terms of order $\O(m_s/m_b)$.
In particular in MSSM, due to the universality of the $A$--terms,
the gluino and chargino contributions
to $\tilde{Q}_{7,8}$ turn out to be of order $\O(m_s/m_b)$.
This argument
does not hold in the models with non--universal $A$--terms and in
particular in our case.
It can be simply understood by using the mass insertion method \cite{GGMS}.
For instance, the gluino contributions to $Q_7$ and $\tilde{Q}_7$
operators are proportional to $(\delta_{LR}^d)_{23}\simeq
(S_{D_L}Y^{A\star}_d S_{D_R}^{\dag})_{23}/m_{\tilde{q}}^2$
and $(\delta_{RL}^d)_{23}\simeq
(S_{D_R}Y^{A}_d S_{D_L}^{\dag})_{23}/m_{\tilde{q}}^2$ respectively.
Since the $A^D$ matrix is symmetric in model A and
$A^D_{ij}\simeq A^D_{ji}$ in model B, then
$(\delta_{LR}^d)_{23}\simeq (\delta_{RL}^d)_{23}$.
Then in our case we should
consistently take into account the SUSY contributions to $\tilde{Q}_{7}$ in
\bsg.
Analogous considerations hold for the operator $\tilde{Q}_{8}$.

By taking into account the above considerations regarding the
operators $\tilde{Q}_{7,8}$,
the new physics effects in \bsg
can be parametrized in a model independent way
by introducing the so called $R_{7,8}$ and $\tilde{R}_{7,8}$
parameters defined at EW scale as
\be
R_{7,8}=\frac{\left(C^{(0)}_{7,8}-C^{(0)SM}_{7,8}\right)}
{C_{7,8}^{(0)SM}},~~~
\tilde{R}_{7,8}=\frac{\tilde{C}_{7,8}^{(0)}}{C_{7,8}^{(0)SM}},
\label{R78}
\ee
where $C_{7,8}$ include the total contribution while $C_{7,8}^{SM}$ contains
only the SM ones.
Note that in $\tilde{C}_{7,8}$, which are the corresponding Wilson coefficients
for $\tilde{Q}_{7,8}$ respectively,
 we have set to zero the SM contribution.
In Ref.~\cite{BBMR} only the expressions for the $R_{7,8}$ are given, 
the corresponding expressions for $\tilde{R}_{7,8}$ are given in Ref.~\cite{emidio}.
The general parametrization of the branching ratio in terms
of the new physics contributions is given by \cite{gabsarid}. 
\bea
\hspace{-2.5cm}BR(B\to X_s\gamma)&=&(3.29\pm 0.33)\times 10^{-4}
\biggl(1 + 0.622 Re[R_7] + 0.090\bigl(\vert R_7 \vert^2 + \vert \tilde{R}_7 
\vert^2\bigr) +\no
&&\hspace{-1cm}0.066Re[R_8]+ 0.019 \bigl(Re[R_7 R_8^*] + Re[\tilde{R}_7 \tilde{R}_8^*]\bigr) + 
0.002 \bigl(\vert R_8 \vert^2+\vert \tilde{R}_8 \vert^2\bigr)\biggr),
\label{bsgPAR}
\eea
where the overall SM uncertainty has been factorized outside.
We have checked explicitly that the result in Eq.(\ref{bsgPAR}) is
in agreement with the corresponding one used in Ref.~\cite{DTV}.
\begin{figure}[htcb]
\centerline{
\begin{tabular}{c}
\epsfig{figure=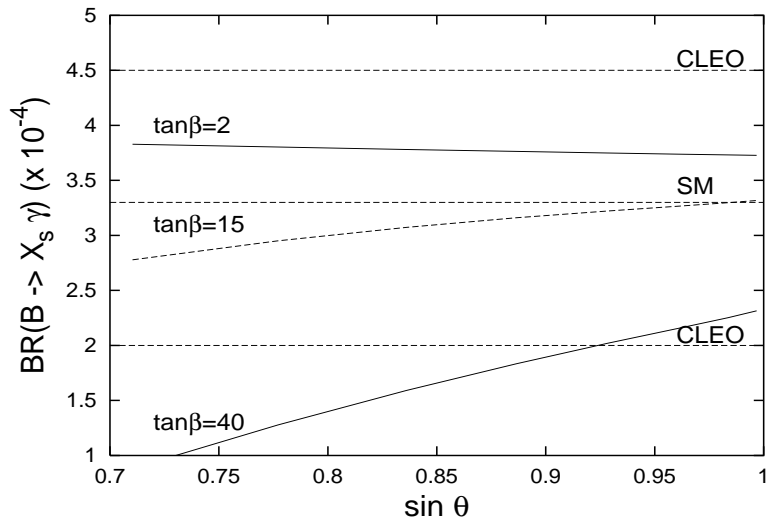,height=7cm,width=10.5cm}\\
\end{tabular}
}
\caption{The BR($\Bsg$) versus $\sin\theta$ in model A,
for $\mu>0$, $m_{3/2}=150$ GeV and \tgb=2, 15, 40.}
\label{BRA1}
\end{figure}

In Figs. [\ref{BRA1}] we plot the results
for the branching ratio BR($\Bsg$), in model A,  versus $\sin{\theta}$
for different values of \tgb, $m_{3/2}=150$ and  $\mu >0$ 
(for $\mu <0$, as in MSSM, 
almost the whole range of the parameter space is excluded).
The main message arising from these results
is that the sensitivity of BR($\Bsg$) respect to $\sin \theta$
increases with \tgb. In particular for the low \tgb region the \bsg result
does not differ significantly from the universal case. In the large
\tgb region, $\tgb=15-40$, the CLEO measurement of \bsg set severe
constraints on the angle  $\theta$ for low gravitino masses.

\begin{figure}[tbc]
\centerline{
\begin{tabular}{c}
\epsfig{figure=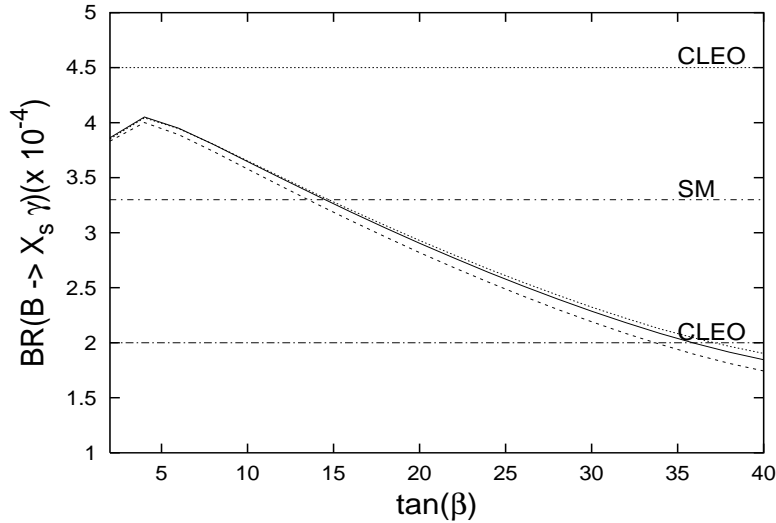,height=7cm,width=10.5cm,angle=0}\\
\end{tabular}
}
\caption{The branching ratio BR($\Bsg$) versus
$\tan \beta$ in model B, for $\mu > 0$, $m_{3/2}=150$ GeV,
and for some values of $(\Theta_1,\Theta_2)=
(1/\sqrt{3},1/\sqrt{3}),~(0.9,0.2),~(0.6,0.2)$, corresponding to the
continuos, dashed, and dot-dashed lines respectively.}
\label{BRB1}
\end{figure} 

In Fig.[\ref{BRB1}] we
plot the branching ratio BR($\Bsg$), in model B, versus \tgb for three different
values of $\Theta_1,\Theta_2$ (see  the figure caption)
which are representative examples for universal and highly non--universal cases.
From these figures it is clear that BR($\Bsg$) is not very sensitive to the
values of $\Theta_i$'s parameters, even at very large \tgb, unlike model A.
The constraints from CLEO measurement are almost the same in the
universal and non--universal cases. For $\mu > 0$
the branching ratio is constrained from the lower bound of CLEO only
at very large \tgb, while for $\mu < 0$ the branching ratio is
almost excluded except at low \tgb.

\section{SUSY phases and CP asymmetry in $B \to
X_s \gamma$ decays}

Direct CP asymmetry in the inclusive radiative decay $B \to X_s \gamma$ is measured
by the quantity
\bea
A_{CP}^{b\to s \gamma} = \Frac{\Gamma(\bar{B} \to X_s \gamma) - \Gamma(B \to
X_{\bar{s}}
\gamma)}{\Gamma(\bar{B} \to X_s \gamma) + \Gamma(B \to X_{\bar{s}}
\gamma)}.
\eea
The Standard Model (SM) prediction for this asymmetry is very small, less than
$1\%$. Thus, the observation of sizeable asymmetry in the decay $B \to X_s \gamma$
would be a clean signal of new physics.

The most recent result reported by CLEO collaboration for the CP asummetry in  these
decays is \cite{cleo2}
\begin{equation}
-9\% < A_{CP}^{b\to s \gamma}  < 42 \%~,
\end{equation}
and it is expected that the measurements of $A_{CP}^{b\to s \gamma}$ will be
improved in the next few years at the $B$--factories.

Supersymmetric predictions for $A_{CP}^{b\to s \gamma}$ are strongly dependent
on the flavour structure of the soft breaking terms. It was shown that in the
universal case, 
as in the minimal supergravity models, the prediction
of the asymmetry is less than $2\%$, since in this case the electric dipole moments
(EDM) of the electron and neutron constrain the SUSY CP--violating phases to be very
small~\cite{Goto,Aoki}. we explore the effect of these large flavour--dependent 
phases on inducing a direct CP violation in $B \to X_s \gamma$ decay. We will show 
that the values of the asymmetry $A_{CP}^{b\to s \gamma}$ in this class of models 
are much
larger than the SM prediction in a wide region of the parameter space allowed by
experiments, namely the EDM experimental limits and the bounds on the branching 
ratio
of $B \to X_s \gamma$ . The enhancement of $A_{CP}^{b\to s \gamma}$ is due to the  
important contributions from gluino--mediated diagrams, in this
scenario, in addition to the usual chargino and charged Higgs contributions.

The expression for the asymmetry  $A_{CP}^{b\to s \gamma}$, corrected to
next--to--leading order is given by \cite{Neubert}
\bea
A_{CP}^{b\to s \gamma} &=&\Frac{4\al_s(m_b)}{9 \vert C_7 \vert^2} \biggl\{   
\biggl[\frac{10}{9} - 2 z~ (v(z)+b(z,\delta))\biggr] Im[C_2 C_7^*]\no
&+& Im[C_7 C^*_8] + \frac{2}{3} z~b(z,\delta) Im[C_2 C_8^*] \biggr \},
\label{asymmetry}
\eea
where $z=m_c^2/m_b^2$. The functions $v(z)$ and $b(z,\delta)$ can be found in
Ref.\cite{Neubert}. The parameter $\delta$ is related to the experimental cut
on the photon
energy, $E_{\gamma} > (1-\delta) m_b/2$, which is assumed to be 0.9. We neglect the
very small effect of the CP--violating phase in the CKM matrix.
As mentioned above, SUSY models with non--universal $A$--terms may induce
non--negligible contributions
to the dipole operators $\tilde{Q}_{7,8}$ which have opposite chirality to $Q_{7,8}$.
In the MSSM these contributions are suppressed by terms of order ${\cal O}(m_s/m_b)$
due to the universality of the $A$--terms. However, in our case we should take them into
account. Denoting by $\tilde{C}_{7,8}$ the Wilson coefficients multiplying the new
operators $\tilde{Q}_{7,8}$ the expression for the asymmetry in Eq.(\ref{asymmetry})
will be modified by making the replacement
\be
C_i C_j^* \to C_i C_j^* + \tilde{C}_i \tilde{C}_j^*.
\label{chirality}
\ee
The expressions for $\tilde{C}_{7,8}$ are given in
Ref.\cite{emidio} and $\tilde{C}_2=0$ (there is no operator similar to $Q_2$
containing right--handed quark fields).

Note that including these modifications (\ref{chirality})
may enhance the branching ratio of $B \to X_s \gamma$ and reduce the CP asymmetry,
since $\vert C_7 \vert^2$ is replaced by $\vert C_7 \vert^2 + \vert \tilde{C}_7
\vert^2$ in the denominator of Eq.(\ref{asymmetry}). If so, neglecting this
contribution could lead to an incorrect conclusion.

In this class of models we consider here, the relevant and
important phase for the CP asymmetry is the phase of the off--diagonal element 
$A_{23}$ ($\phi_b$).
In Fig.1 we show the dependence of $A_{CP}^{b\to s \gamma}$ on
$\phi_b$ for $m_{3/2} =150$ GeV and $\tan\beta=3$ and $10$. 

\begin{figure}[htcb]
\centerline{
\begin{tabular}{c}
\epsfig{figure=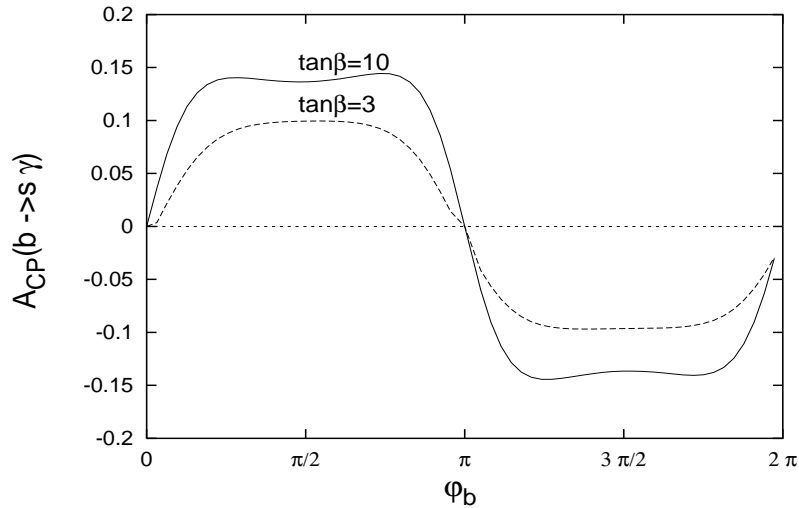,height=7cm,width=10.5cm}\\
\end{tabular}
}
\caption{CP asymmetry $A_{CP}^{b\to s \gamma}$ as a
function of the flavour--dependent phase $\phi_b$, for $m_{3/2} \simeq 150$ GeV and
$\tan \beta =3$ and $10$.}
\label{CPA}
\end{figure}
We see from Fig.[\ref{CPA}] that the CP asymmetry $A_{CP}^{b\to s \gamma}$ can be as
large as $\pm 15\%$, which
can be accessible at the B--factories. Also this result does not require a light chargino
as in the case considered in Ref.~\cite{Ko}.

It is important to emphasize that the gluino contribution in this model gives the
dominant contribution to the CP asymmetry $A_{CP}^{b\to s \gamma}$. We found
that although the real parts of the gluino contributions to both of $C_{7,8}$ and
$\tilde{C}_{7,8}$ are smaller than the real parts of the other
contributions (but not negligible as in the case of universal $A$--terms), their
imaginary parts are dominant and give with the imaginary parts
of the chargino contribution the main contributions to $A_{CP}^{b\to s \gamma}$. It
is clear that these contributions vanish for $\phi_b$ equal to a multiple of $\pi$
and $A_{CP}^{b\to s \gamma}$ in this case is identically zero as Fig.1 shows.
We also noted~\cite{david} that large
values of CP asymmetry $A_{CP}^{b\to s \gamma}$ prefer small values for the 
branching ratio $BR(B\to X_s \gamma)$. This correlation is also found in 
Ref.~\cite{Aoki}. 

\section{Conclusions}
We analysed the constraints set by the \bsg decay
on a class of string inspired SUSY models with non--universal soft breaking $A$--terms. We found that the recent CLEO measurements on the total inclusive
B meson branching ratio BR($\Bsg$) do not set severe constraints on the
non--universality of these models.
In this respect we have found that the parameter regions
which are important for generating sizeable contributions to
$\varepsilon^{\prime}/\varepsilon$ \cite{khalil1,khalil2,vives},
in particular the low \tgb regions, are not excluded by \bsg decay.

we have also considered the possible supersymmetric contribution
to CP asymmetry in the inclusive $B\to X_s \gamma$ decay in model with
non--universal $A$--terms. Contrary to the universal scenario, we
find that the CP asymmetry in this class of models is predicted to be large
in sizeable regions of the parameter space allowed by the experimental bounds,
and may be possibly to be detected at B factories  We have shown that the
flavour--dependent phases
are crucial for this enhancing with respecting the severe bounds on the electric
dipole moment of the neutron and electron.

\section*{ Acknowledgement}
I would like to thank D. Bailin, E. Gabrielli, and E. Torrente-Lujan for their
collaboration in this project. I also would like to thank the organizers for such
a nice and stimulating atmosphere in which the workshop took place.\\




\begin{thebibliography}{99}

\bibitem{CLEO}
S. Ahmed {\it et al.}, (CLEO Collaboration), CLEO-CONF-99-10,
hep-ex/9908022.
\bibitem{ALEPH}
R. Barate {\it et al.} (ALEPH Collaboration), \plb{429}{1998}{169}.
\bibitem{bsgSUSY}
W. S. Hou and R.S. Willey \plb{202}{1988}{591};
T. G. Rizzo \prd{38}{1988}{820};
V. Barger, M.S. Berger, and R.J.N. Phillips, \prl{70}{1993}{1368};
J. L. Hewett, \prl{70}{1993}{1045};
R. Barbieri and G.F. Giudice, \plb{309}{1993}{86};
J. L. Lopez, D. V. Nanopoulos, and G. T. Park, \prd{48}{1993}{974};
Y. Okada, \plb{315}{1993}{119};
R. Garisto and J.N. Ng, \plb{315}{1993}{372};
M.A. Diaz, \plb{322}{1994}{207}; F.M. Borzumati, Z. Phys.
{\bf C 63} (1994) 291;
P. Nath and R. Arnowitt, \plb{336}{1994}{395};
S. Bertolini and F. Vissani, Z. Phys. {\bf C 67} (1995) 513;
N.G. Deshpande, B. Dutta, and S. Oh, \prd{56}{1997}{519};
S. Khalil, A. Masiero, and Q. Shafi, \prd{56}{1997}{5754};
T. Blazek and S. Raby,\prd{59}{1999}{095002}.

\bibitem{BBMR}
S. Bertolini, F. Borzumati, A. Masiero, and G. Ridolfi,  
\npb{353}{1991}{591}. 
\bibitem{barr}
S.~Barr and S.~Khalil, \prd{61}{2000}{035005}.
\bibitem{khalil1}
S.~Khalil, T.~Kobayashi, and A.~Masiero, \prd{60}{1999}{075003}.
\bibitem{khalil2}
S.~Khalil and T.~Kobayashi, \plb{460}{1999}{341}.
\bibitem{vives}
S. Khalil, T. Kobayashi, and O. Vives,  \npb{580}{2000}{275}. 
\bibitem{emidio}
E. Gabrielli, S. Khalil, and E. Torrente-Lujan, hep-ph/0005303. 
\bibitem{david}
D.Bailin and S. Khalil, hep-ph/0010058 . 
\bibitem{ibanez1}
A. Brignole, L. E. \Ibanez, and C. \Munoz, \npb{422}{1994}{125},
Erratum-ibid. {\bf B 436}~(1995) 747.
\bibitem{ibanez2}
L. E. \Ibanez, C. \Munoz, and S. Rigolin, \npb{553}{1999}{43}.
\bibitem{bsgNLO}
K. Chetyrkin, M. Misiak, and M. Munz, Phys. Lett. B {\bf 400}, 206 (1997);
A. J. Buras, A. Kwiatkowski, and N. Pott, Phys. Lett. B {\bf 414}, 157 (1997);
C. Greub and T. Hurth, \prd{54}{1996}{3350}; \prd{56}{1997}{2934}.
\bibitem{GGMS}
F. Gabbiani, E. Gabrielli, A. Masiero, and L. Silvestrini,
\npb{477}{1996}{321}.
\bibitem{gabsarid}
E. Gabrielli and U. Sarid \prd{58}{1998}{115003}; \prl{79}{1997}{4752}.
\bibitem{DTV}
M. A. Diaz, E. Torrente-Lujan, and J.W.F. Valle, \npb{551}{1999}{78}.
\bibitem{cleo2}
CLEO Collaboration, S. Ahmed et al., hep-ex/9908022.
\bibitem{Goto} 
T. Goto, Y. Keum, T. Nihei, Y. Okada and Y. Shimizu, \plb{460}{99}{333};
\bibitem{Aoki}   
M. Aoki, G. Cho and N. Oshimo, \npb{554}{99}{50}.
\bibitem{Neubert}
A. Kagan and M. Neuber, \prd{58}{98}{094012}.
\bibitem{Ko}
S. Baek and P. Ko, \prl{83}{99}{488}.

\end{thebibliography}
\end{document}